\documentclass[aps,prx,twocolumn,showpacs]{revtex4-1}
\usepackage{graphics}
\usepackage{subfigure}
\usepackage{longtable}
\usepackage{graphicx}
\usepackage{pstricks}
\usepackage{amsmath}
\usepackage{dcolumn}
\usepackage{bm}
\usepackage{pdfpages}

\begin{document}

\title{Prospect of quantum anomalous Hall and quantum spin Hall effect in doped kagome lattice Mott insulators}

\author{Daniel Guterding}
\email{guterding@itp.uni-frankfurt.de}
\affiliation{Institut f\"ur Theoretische Physik, Goethe-Universit\"at Frankfurt, 
Max-von-Laue-Stra{\ss}e 1, 60438 Frankfurt am Main, Germany}

\author{Harald O. Jeschke}
\affiliation{Institut f\"ur Theoretische Physik, Goethe-Universit\"at Frankfurt, 
Max-von-Laue-Stra{\ss}e 1, 60438 Frankfurt am Main, Germany}

\author{Roser Valent\'i}
\affiliation{Institut f\"ur Theoretische Physik, Goethe-Universit\"at Frankfurt, 
Max-von-Laue-Stra{\ss}e 1, 60438 Frankfurt am Main, Germany}

\begin{abstract}
Electronic states with non-trivial
topology host a number of novel phenomena with potential for
revolutionizing information technology. The quantum anomalous Hall
effect provides spin-polarized dissipation-free transport of
electrons, while the quantum spin Hall effect in combination with
superconductivity has been proposed as the basis for realizing
decoherence-free quantum computing. We introduce a new strategy for
realizing these effects, namely by hole and electron doping kagome
lattice Mott insulators through, for instance, chemical substitution.
As an example, we apply this new approach to the natural mineral
herbertsmithite. We prove the feasibility of the proposed
modifications by performing \textit{ab-initio} density functional
theory calculations and demonstrate the occurrence of the predicted
effects using realistic models. Our results herald a new family of
quantum anomalous Hall and quantum spin Hall insulators at affordable
energy/temperature scales based on kagome lattices of transition metal
ions.
\end{abstract}


\maketitle

\section*{Introduction}
\begin{figure*}[t]
\includegraphics[width=0.75\linewidth]{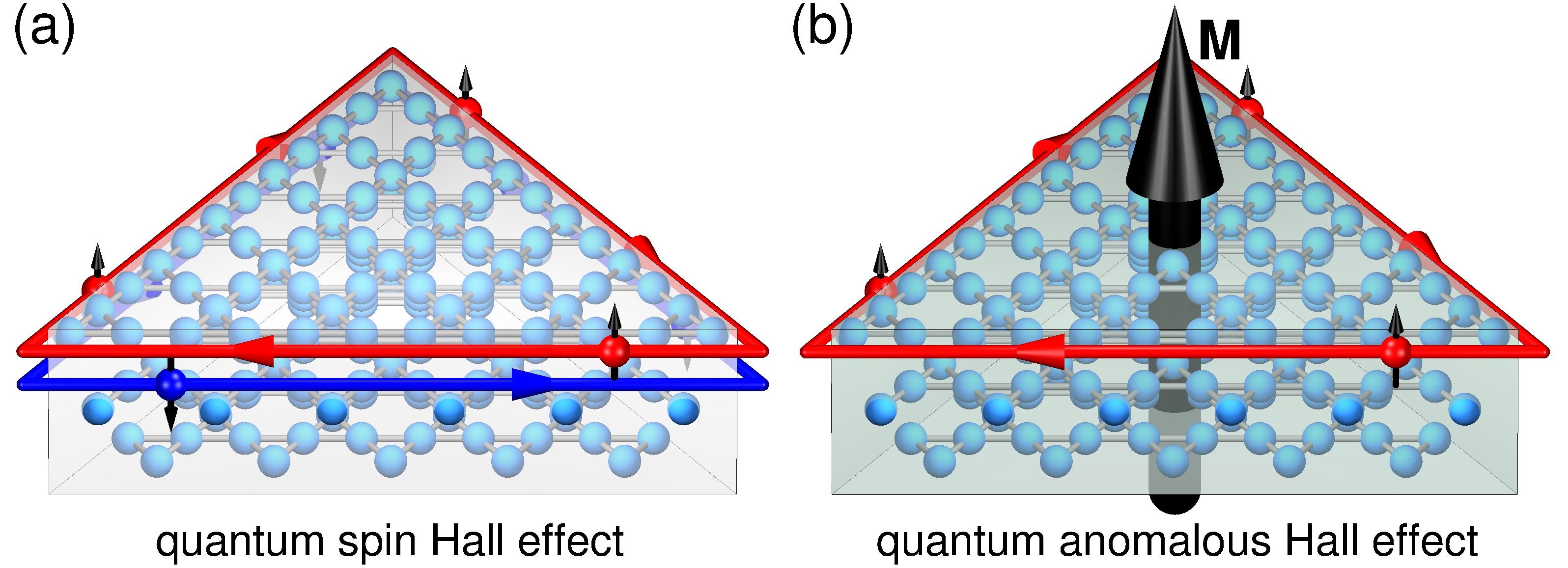}
\caption{ Predicted quantum Hall effects in doped
    herbertsmithite. (a) Quantum spin Hall effect (QSHE)
  with counterpropagating edge currents of opposite spin-polarization.
  (b) Quantum anomalous Hall effect (QAHE) with a single
  spin-polarized edge current due to intrinsic magnetization.  In both
  cases the crystal is cut into triangular shape. The crystal
  structure shows the arrangement of copper atoms (light blue balls) in
  three shifted kagome planes as in herbertsmithite.  }
\label{fig:halleffects}
\end{figure*}

The kagome lattice structure, which consists of corner-sharing
triangles, is notorious for supporting exotic states of matter. For
instance, the possible experimental realization of quantum
spin-liquids based on spin-1/2 kagome lattices has generated in the
past intense research efforts on herbertsmithite and similar
frustrated antiferromagnets~\cite{HerbertsmithiteSynthesis,
  LeeDrought, BalentsQSL, Mendels2010, HerbertsmithiteExchange,
  Colman2010, Li2014, Han2012, BauerAnyons, AntisiteDisorder,
  AntisiteDisorder2, GaHerbertsmithite}.
Recently, the kagome lattice has also received plenty of attention for
quasiparticle excitations with non-trivial
topology~\cite{MookEdgeStates, PereiroSkyrmions,
  TopologicalMagnonsChisnell}. From topologically non-trivial
electronic bands, effects such as the quantum spin Hall effect
(QSHE)~\cite{KaneMeleModel, BernevigZhangModel, HasanKaneColloquium}
and the quantum anomalous Hall effect
(QAHE)~\cite{HaldaneModel,Liu2016} can emerge, also in kagome
lattices~\cite{KagomeQSHEWang, KagomeIntrinsic}.
A quantum spin Hall insulator in two dimensions, also known as a
topological insulator, is a topological state of matter, present in a
system with spin-orbit coupling, where symmetry protected
dissipationless spin-polarized currents counterpropagate on the sample
edges, while the bulk of the sample remains insulating
(Fig.~\ref{fig:halleffects}a).  This phenomenon has received
considerable attention because Majorana bound states have been
predicted to appear at interfaces between QSHE materials and
superconductors~\cite{FuKaneMajorana, WilczekMajorana,
  HasanKaneColloquium}.  Employing these Majorana zero modes for
topological quantum computation is a rapidly developing
field~\cite{DasSarmaTopologicalComputing}.

In contrast to the QSHE, in a quantum anomalous Hall insulator only
one spin species propagates around the sample edge due to the presence
of intrinsic magnetization in the sample (Fig.~\ref{fig:halleffects}b).
This state of matter offers a direct realization of {\it
  intrinsic} topological properties in a material through the
combination of spin-orbit coupling and
magnetism~\cite{BernevigZhangModel}.  Due to the dissipation-free,
spin-polarized edge currents in the absence of external magnetic
fields, realizations of the QAHE are also intensively sought for,
especially for application in new energy-efficient spintronic
devices~\cite{He2015,Weng2015}.  So far, in electronic systems the
QAHE has only been observed in thin films of chromium-doped
(Bi,Sb)$_2$Te$_3$ at a temperature of
$30~\mathrm{mK}$~\cite{QAHEMagneticTITheory,
  QAHEMagneticTIExperiment}, the main limitation being the low Curie
temperature of the material involved. Lately, it has been proposed
that the QAHE can be realized in some other compounds using, for
instance, manipulated surfaces or exfoliated
monolayers~\cite{ZhangManganeseKagome, YanQAHEHoneycomb,
  AdatomsVanderbilt}.  Another interesting approach is design from
scratch of organometallic networks with topological
bandstructures~\cite{LiuOrganometallic, AokiOrganometallic}. 
A good
strategy for designing QAHE compounds based on existing materials with
favorable energy scales that are adequate for applications is however
currently lacking.

\begin{figure}[b]
\includegraphics[width=\linewidth]{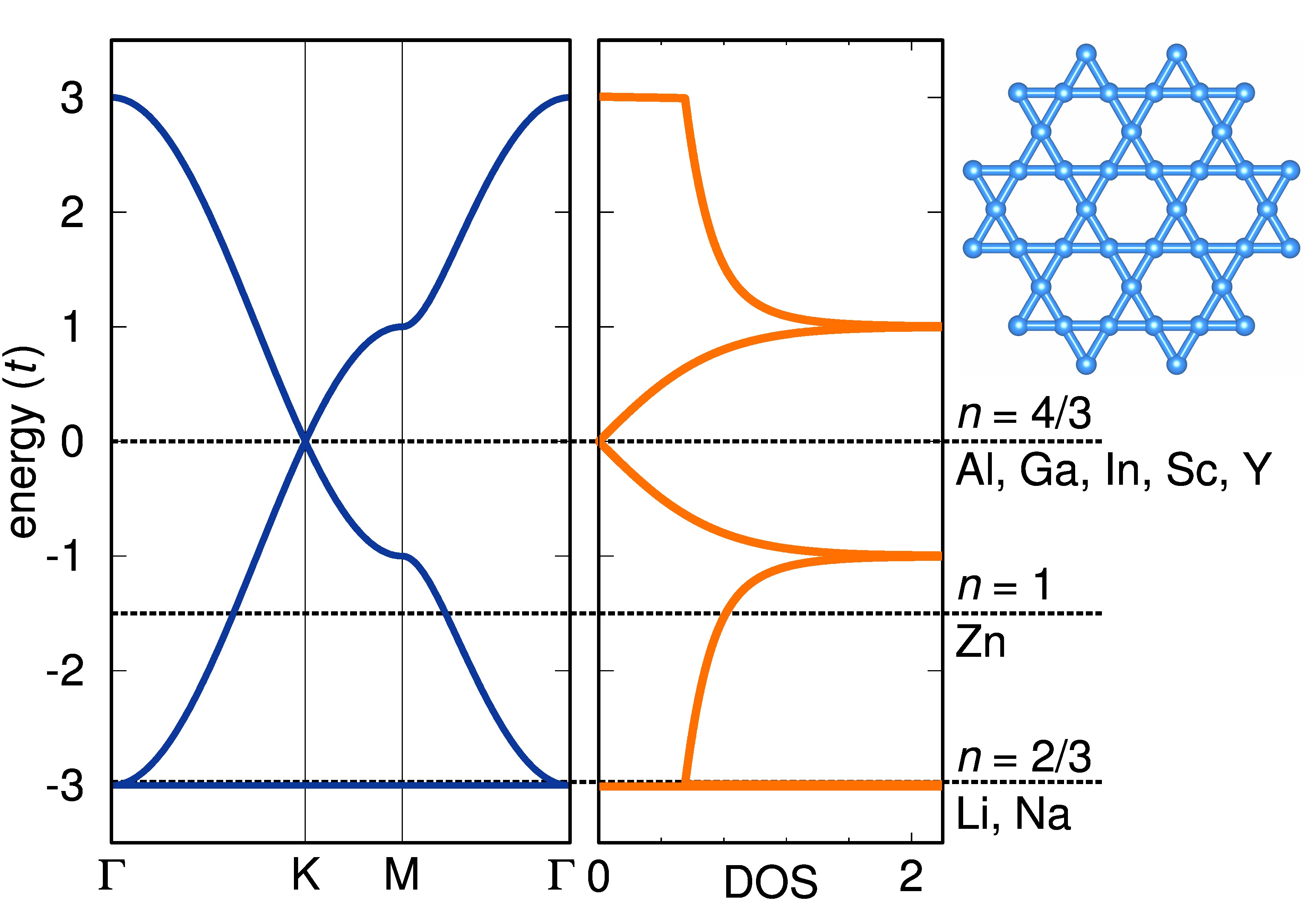}
\caption{Electronic bandstructure and density of states of the
    pure kagome lattice (shown in upper right corner). On the right hand side the possible
  substituent elements for the herbertsmithite system with
  corresponding Fermi level are listed.}
\label{fig:purekagomedosfilling}
\end{figure}

Here, we propose a new approach to create materials with non-trivial band 
topology and large Curie temperatures, exploiting the electronic properties of 
doped Mott insulators on a kagome lattice.  A quick look at the one-electron 
properties (bandstructure) of the kagome lattice with nearest neighbor hoppings 
(Fig.~\ref{fig:purekagomedosfilling}) shows huge potential for the realization 
of possible exotic states by only varying the electron filling. At half-filling 
the Fermi level lies near a van Hove singularity and inclusion of many-body 
correlation effects renders the system a Mott insulator~\cite{KagomeDMFT}. 
At a filling of $n=4/3$, however, the Mott transition is absent~\cite{GaHerbertsmithite} 
and the Fermi level is at the Dirac point, where non-trivial band effects may be 
expected upon consideration of spin-orbit coupling. The spin-orbit coupling 
opens a gap at the position of the Dirac point and the non-trivial topology of 
electrons on the kagome lattice leads to surface states of both spin species 
that traverse the bulk band gap opened by relativistic effects and, the QSHE is 
realized.

Even more interesting is the filling of $n=2/3$ with the Fermi level
right at the flat band.  It was recently
suggested~\cite{WenHighTempFQHE} that if a nearly flat band is
partially filled, a proper combination of spin-orbit coupling,
ferromagnetism and geometric frustration will give rise to the
fractional quantum Hall effect at high temperatures.  Along these
lines, we exploit here as a key ingredient for topological non-trivial
states, the tendency towards ferromagnetism~\cite{Hanisch1997} of a
filled flat band in hole-doped transition-metal-based kagome lattices.
At $n=2/3$ the ferromagnetic instability combined with correlation
effects is expected to gap out one spin-channel and move the Fermi
level of the other spin-channel exactly to the Dirac point. When
spin-orbit coupling (SOC) is considered, we have the same situation as
for the filling of $n=4/3$ but only for one spin species. In such a
situation, the QAHE with fully spin-polarized dissipation-free surface
states is realized.

To demonstrate this new strategy of finding QSHE and QAHE materials by
doping Mott insulators, we investigate which possible modifications of
the natural mineral herbertsmithite -a Mott insulator with spin-liquid
behavior- leave the perfect kagome motif undistorted and realize
different electronic fillings.

\begin{figure*}[t]
\includegraphics[width=0.65\linewidth]{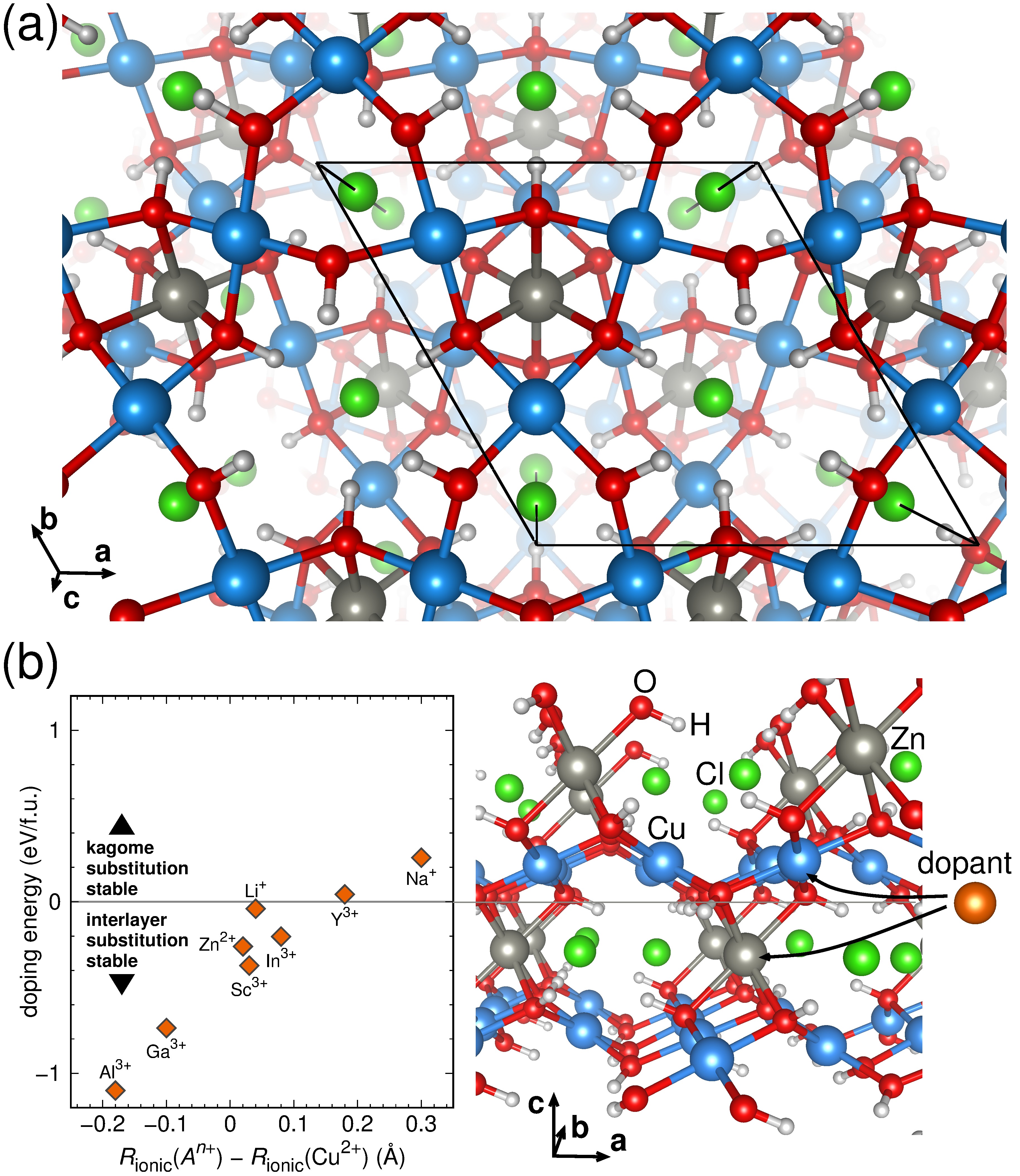}
\caption{ Crystal structure and calculated doping energies for
  herbertsmithite. (a) View of the crystal structure of
  herbertsmithite along the c-axis, which is perpendicular to the
  copper kagome layers. The Cu atoms (shown in light blue) are in a square planar crystal
  field environment of oxygen ions (red) so that the orbitals near the Fermi
  level are correlated $d_{x^2-y^2}$ states. Each oxygen atom bonds to
  a hydrogen atom (white) located outside the copper and oxygen layer. In this
  interlayer space also zinc (grey) and chlorine (green) atoms are located.
  (b) The left hand side shows doping
  energies for herbertsmithite. All data points above the zero energy
  line indicate that the kagome lattice will be distorted upon
  doping. The right hand side shows the crystal structure of
  herbertsmithite with two possible sites for substitution
  indicated. On the interlayer site (Zn position, negative doping
  energy), herbertsmithite mostly prefers to incorporate ions with
  smaller radius than Cu$^{2+}$. All ions with positive doping energy
  occupy the Cu position and distort the kagome lattice. Ionic radii
  in coordination number 6 are taken from Ref.~\cite{IonicRadii}.}
\label{fig:dopingenergies}
\end{figure*}

Herbertsmithite crystallizes in the centrosymmetric space group
$R\,\bar{3}m$ and its structure is based on layers of Cu$^{2+}$
($3d^9$) ions building a perfect two-dimensional half-filled
frustrated kagome lattice separated by layers of Zn$^{2+}$ ions
(Fig.~\ref{fig:dopingenergies}a). The Cu atoms are in a square planar crystal
field environment of oxygen ions so that the orbitals near the Fermi
level are correlated $d_{x^2-y^2}$ states.  Evaluating density
functional theory (DFT) total energies, we show that single crystals of
materials obtained by following various doping choices in
herbertsmithite can in principle be synthesized. Further, we prove
that the magnetic ground state of hole-doped herbertsmithite at
filling $2/3$ is ferromagnetic, which validates that the flat band
physics of the pure kagome lattice carries over to realistic
situations. Finally, we demonstrate the presence of topologically
non-trivial surface states of doped herbertsmithite using a
state-of-the-art Wannier function technique based on fully
relativistic DFT calculations.

\section*{Materials and Methods}
We prepared hypothetical materials starting from the experimental
crystal structure of herbertsmithite~\cite{HerbertsmithiteSynthesis},
substituting zinc (Zn$^{2+}$) atoms between the copper kagome layers
(see Fig.~\ref{fig:dopingenergies}) by monovalent
$A$=Li$^{+}$,~Na$^{+}$ (hole-doping) and trivalent
Al$^{3+}$,~Ga$^{3+}$,~In$^{3+}$,~Sc$^{3+}$,~Y$^{3+}$
(electron-doping). We refer to these compounds as $A$-herbertsmithite,
$A$Cu$_3$(OH)$_6$Cl$_2$.

Experimental and hypothetical crystal structures were fully relaxed
using DFT in the projector augmented wave (PAW) formulation~\cite{PAWmethod}
implemented in {\sc GPAW}~\cite{GPAWmethod} with a plane-wave cutoff
of $1000~\mathrm{eV}$ and the GGA exchange-correlation
functional~\cite{PerdewBurkeErnzerhof}. We optimized the
stoichiometric structures using $6^3$ $k$-points ($4^3$ $k$-points for
non-stoichiometric structures) until forces were below $10\,
\mathrm{meV/\AA}$.

For each of the substituted structures with perfect copper kagome
layer we also constructed a defect structure, where we lowered the
symmetry of the unit cell and exchanged the substituent $A$ with a
copper atom from a kagome lattice site.  As the chemical composition
of these defect structures is identical to the defect-free structures,
energy differences can be evaluated directly within DFT.  In case the
defect structure has lower energy, the kagome lattice is likely to be
destroyed and the phenomena of interest here will not arise in the
target compound.

Total energies, electronic bandstructures and magnetic exchange
interactions of the relaxed structures were then evaluated using
{\it ab-initio} DFT calculations within an all-electron full-potential 
local orbital (\textsc{FPLO})~\cite{FPLOmethod} basis. For the
exchange-correlation functional we employed the generalized gradient
approximation (GGA)~\cite{PerdewBurkeErnzerhof}, as well as
DFT+U~\cite{Liechtenstein95} functionals. The latter was necessary in
order to treat the correlated nature of Cu 3$d$ orbitals.  The Hubbard
repulsion on the Cu 3$d$ orbitals was set to $U=6~\mathrm{eV}$ and
Hund's rule coupling to $J_H = 1~\mathrm{eV}$. Although we concentrate
our investigation on the Cu $d_{x^2-y^2}$ orbitals close to the Fermi level,
the interactions were included in the entire Cu 3$d$ shell, which is spread 
out over a large range of energies due to the interaction with the ligands.
Additionally, we
investigated the effect of spin-orbit coupling on the electronic
bandstructure employing the fully relativistic version of the
\textsc{FPLO} code. Total energies, electronic bandstructures,
tight-binding and Heisenberg models were extracted from
calculations converged using $8^3$, $20^3$, $40^3$ and $6^3$ $k$-point
grids respectively.

To demonstrate the existence of surface states, we constructed bulk
tight-binding models for the copper states $(n, j, m_j) = (3, 5/2, \pm
5/2)$ from fully relativistic DFT calculations using projective Wannier
functions~\cite{FPLOtightbinding}. Employing a method based on Green's
functions~\cite{SemiInfiniteGreensFunctionMethod1,
  SemiInfiniteGreensFunctionMethod2,
  SemiInfiniteGreensFunctionMethod3}, we calculate the states on the
surface of herbertsmithite. The spectral function is obtained from the Green's
function as $A(k,
\omega) = - \text{Im}~ G(k, \omega)/ \pi$.

\section*{Results}
\subsection*{Stability estimates}
By performing exhaustive DFT calculations we identified as the limiting factor 
for modifying herbertsmithite the tendency of certain ions towards substituting 
copper sites in the kagome layer. In Fig.~\ref{fig:dopingenergies}b we plot the 
energy difference (tabulated in the Supplemental Information) between 
substitution at the kagome site and substitution at the interlayer site for 
herbertsmithite as a function of the substituent ionic radius~\cite{IonicRadii}. 
In herbertsmithite, sodium (Na$^{+}$) and yttrium (Y$^{3+}$) prefer to occupy a 
site in the kagome layer, which generates a monoclinically distorted crystal 
structure with no perfect kagome lattice. If the substituent atom occupies the 
interlayer site, the perfect kagome motif is preserved.

In terms of substitution energies, lithium (Li$^{+}$) is the most
promising candidate for synthesis of hole-doped herbertsmithite. On
the electron-doped side, aluminum (Al$^{3+}$), gallium (Ga$^{3+}$) and
scandium (Sc$^{3+}$) are the most promising candidates for
substitution. Formation of the substituted materials is found to be
energetically favorable compared to the formation of the parent
compound clinoatacamite, Cu$_2$(OH)$_3$Cl. All herbertsmithite-based
materials investigated are stable against formation of vacancies and
copper impurities, as opposed to full substitution, on the interlayer
site. We also investigated fractional substitution of Zn$^{2+}$ by Ga$^{3+}$ and
found that the doping series Ga$_x$Zn$_{1-x}$Cu$_3$(OH)$_6$Cl$_2$
should be stable in a broad range of Ga:Zn ratios (see Supplemental Information).

\subsection*{Electronic and magnetic properties}
\begin{figure}[t]
\includegraphics[width=\linewidth]{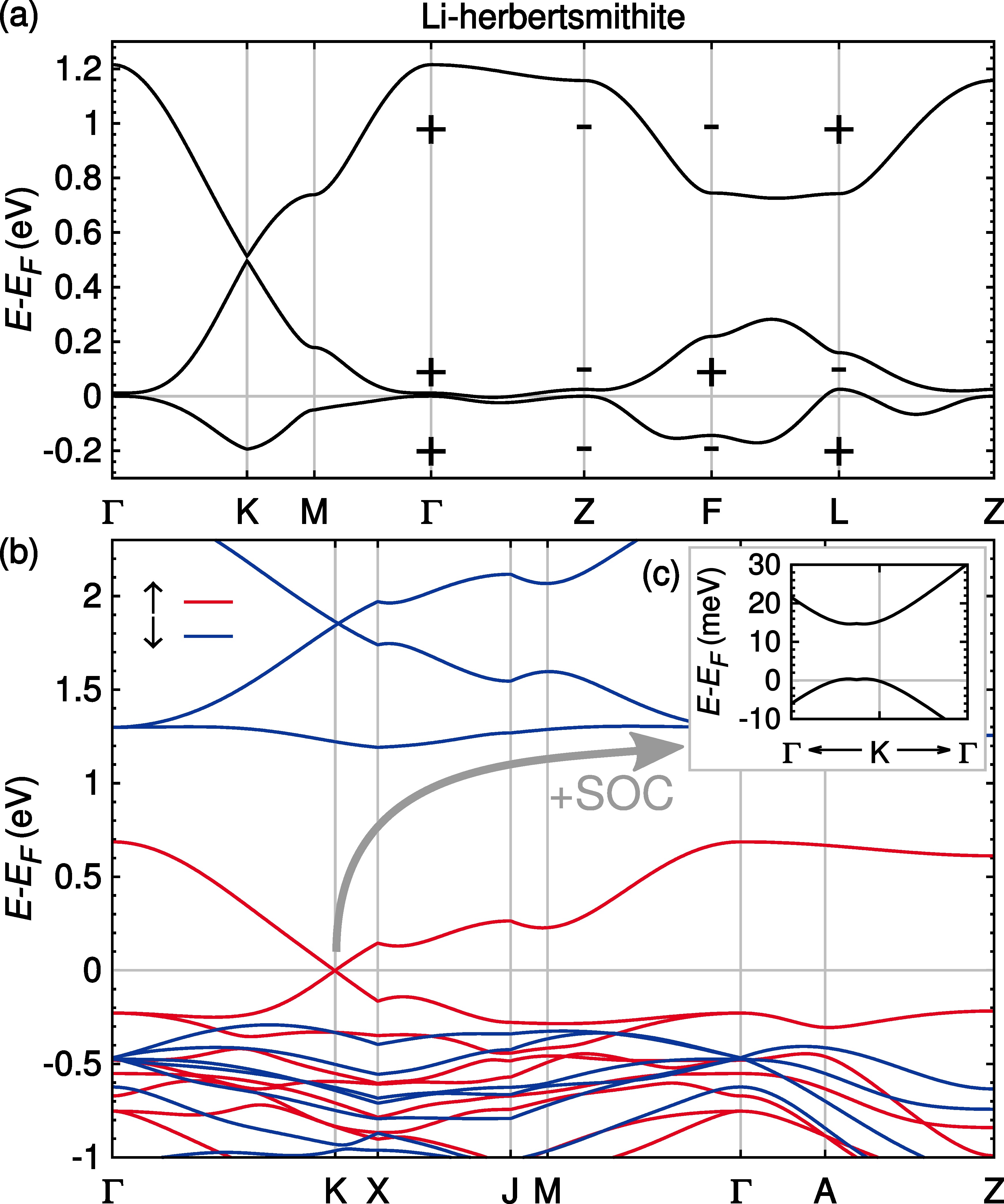}
\caption{(a)
  Fully relativistic non-spin-polarized bandstructure. Plus and minus
  signs denote the parities of the three bands closest to the Fermi
  level at eight inversion-symmetric points ($\Gamma$, $3 \times F$,
  $3 \times L$, $Z$) in the Brillouin zone. (b) FM
  bandstructure without SOC.  The spin down channel is gapped, while
  the Dirac point of the spin up channel is located exactly at the
  Fermi level. The path through the Brillouin zone is shown in
  the Supplemental Information. (c) Fully relativistic FM bandstructure close to
  the Dirac point. SOC opens a gap.}
\label{fig:holedopedherbertsmithitebands}
\end{figure}

In the analysis of the electronic and magnetic properties we
concentrate here on the hole-doped materials ($n=2/3$, as defined in
Fig.~\ref{fig:purekagomedosfilling}) where the Fermi level lies in the
region of the flat band of the kagome lattice and a strong
ferromagnetic instability is to be expected.
Fig.~\ref{fig:holedopedherbertsmithitebands}a displays the
fully relativistic non-spin-polarized electronic bandstructure of
Li-herbertsmithite, where the parities of the three (dominantly Cu $d_{x^2-y^2}$)
bands closest to the Fermi level are also indicated. Due to the bulk
nature of the system, the ideal kagome flat band acquires some
dispersion. Nonetheless, ferromagnetism is strongly favored. Indeed,
total energy calculations of Li-herbertsmithite in various copper spin
configurations give ferromagnetism as the ground state.  Furthermore,
parametrizing the Cu-Cu interactions by mapping DFT+U total energies
to a spin-$1/2$ Heisenberg Hamiltonian shows a large ferromagnetic
nearest neighbor exchange $J_1 = 544~\mathrm{K}$ and other couplings
with magnitude smaller than $0.1 J_1$. Using a
mean-field approximation~\cite{BlundellMagnetism},
we estimate a Curie temperature of $T_C \approx 1160~\mathrm{K}$ 
(see Supplemental Information for further details).

Fig.~\ref{fig:holedopedherbertsmithitebands}b displays the
non-relativistic ferromagnetic bandstructure obtained with the DFT+U
functional. The Fermi level of the majority bands lies right at the
Dirac point
and inclusion of spin orbit coupling opens
a gap of about $15~\mathrm{meV}$ (see inset in
Fig.~\ref{fig:holedopedherbertsmithitebands}). 
The Dirac point position is slightly displaced from $K$ due to
the finite coupling between the kagome layers, as has been observed previously
in Ref.~\onlinecite{GaHerbertsmithite}. 
The position of the
Fermi level is understood by comparing the results to the
spin-degenerate bandstructure shown in
Fig.~\ref{fig:purekagomedosfilling}.
The fully spin-polarized ferromagnetic state occurs at $n=2/3$,
therefore the spin-resolved fillings are $n_\uparrow = 2/3$ and
$n_\downarrow = 0$. As a consequence, the bands of the majority spins
resemble the non-spin-polarized case at $n=4/3$ (where $n_\uparrow =
n_\downarrow = 2/3$) and the minority spins are empty and gapped. The
ferromagnetic instability places therefore the Fermi level of the up
spin bands right at the Dirac point.

To show the presence of topologically protected edge states in doped
herbertsmithite, we calculated the product of parity eigenvalues at
eight inversion-symmetric points (see
Fig.~\ref{fig:holedopedherbertsmithitebands}) in the Brillouin zone
($\Gamma$, $3 \times F$, $3 \times L$, $Z$)~\cite{TIInvariants}. For
all materials of the herbertsmithite family, topological numbers of
the bands below the Dirac point are $\nu_0;(\nu_1, \nu_2, \nu_3) =
0;(111)$. These indices indicate that the system realizes a stack of
two-dimensional topological insulators (so-called {\it weak} TI),
which displays conducting states on a (001)
surface~\cite{TIInvariants}, although the bulk bands are gapped by
relativistic effects.

Note that non-trivial band topology is intrinsic to the perfect kagome
lattice~\cite{KagomeQSHEWang, KagomeIntrinsic} and no particular
inversion of orbital weights is required unlike in most topological
insulators~\cite{TMHalide, QAHEMagneticTITheory}. In real materials
however, the kagome layer is embedded into a crystal, where
non-trivial band-topology can be destroyed by additional
hybridizations. We observed this case for instance in test
calculations for modifications of the natural mineral barlowite,
$A$Cu$_3$(OH)$_6$FBr~\cite{Han2014, BarlowiteTheoryExperiment}, which
has a crystal structure similar to herbertsmithite with perfect kagome
layers.

\subsection*{Demonstration of surface states}

\begin{figure}[t]
\includegraphics[width=\linewidth]{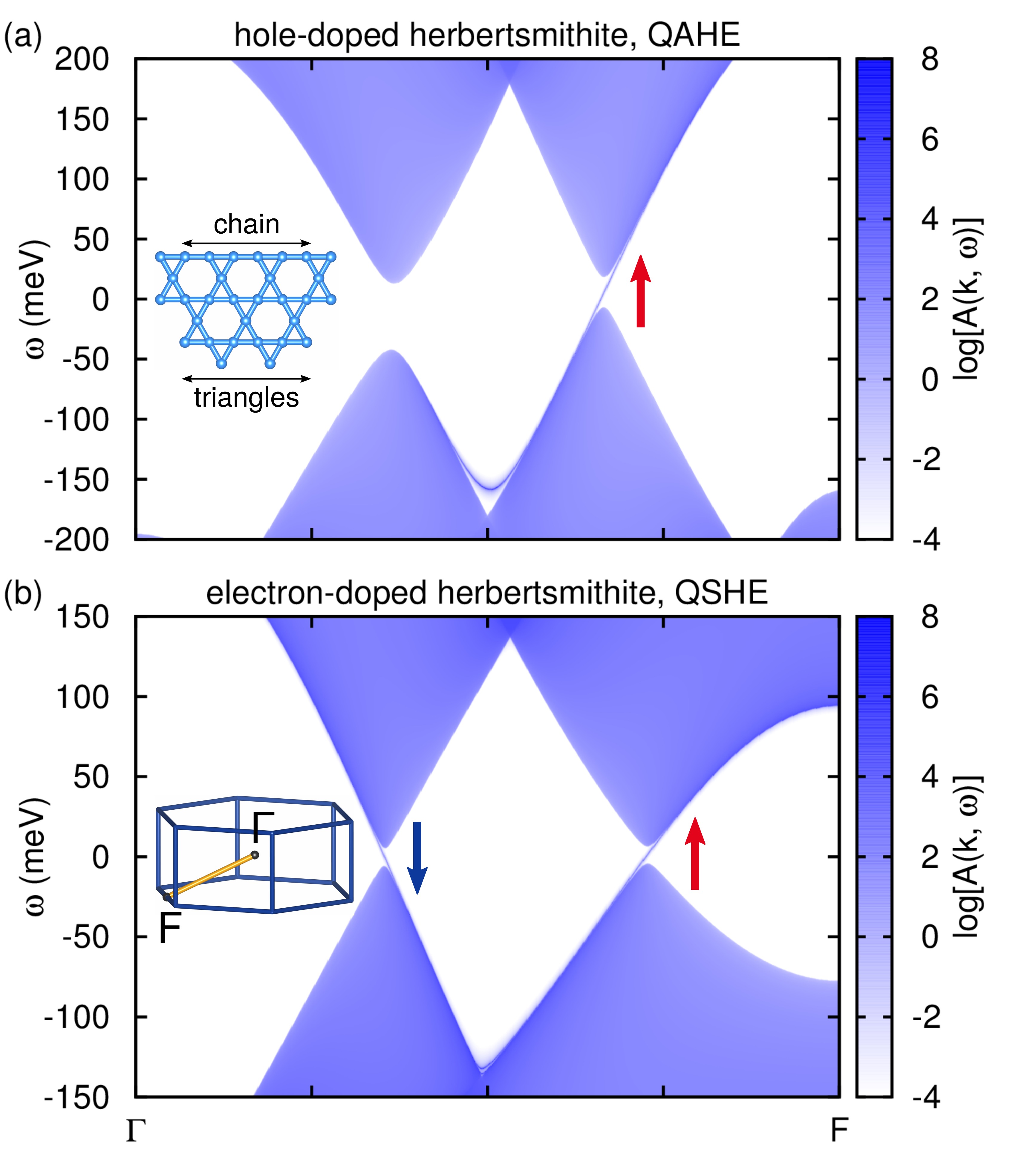}
\caption{Calculated surface states of substituted
    herbertsmithite. Spectral function on the (001) surface of
  (a) Li-herbertsmithite (hole-doped) and (b)
  Ga-herbertsmithite (electron-doped) calculated using Green's
  functions for the semi-infinite system. The arrows pointing upwards / downwards stand
  for the $m_j=+5/2$ and $-5/2$ states respectively. The inset of (a)
  shows the two possible terminations of the kagome lattice. The inset
  of (b) shows the path in the hexagonal Brillouin zone.}
\label{fig:surfacestates}
\end{figure}

Having found non-trivial band-topology in the herbertsmithite system,
we predict that hole-doped herbertsmithite shows a QAHE at $n = 2/3$
filling while electron-doped herbertsmithite shows a QSHE at $n = 4/3$
filling. For both cases we constructed realistic tight-binding models
for the orbitals close to the Fermi energy and calculated the states
on the (001) plane of semi-infinite interlayer-substituted
herbertsmithite (for further details see Supplemental Information). The
obtained spectral function $A(k, \omega)$ of the surface layer in
chain termination is shown in Fig.~\ref{fig:surfacestates}, where the $k$-path
is chosen perpendicular to the direction
in which surface states propagate.

The hole-doped case clearly shows only one surface state of one spin
species crossing the Fermi level (QAHE, see
Fig.~\ref{fig:surfacestates}a), while the electron-doped case
shows two surface states with opposite spin (QSHE, see
Fig.~\ref{fig:surfacestates}b). The spectral function of the
dual surface (triangles termination) has the same essential features
(shown in the Supplemental Information). As we take into account
realistic bandstructures, our spectral functions show additional
structure away from the Fermi level compared to model calculations in
next-neighbor approximation~\cite{KagomeQSHEWang, KagomeIntrinsic}.

\section*{Discussion}
In this work we have presented a new generally applicable strategy for creating 
materials where the quantum anomalous Hall effect and the quantum spin Hall 
effect can be realized at affordable energy and temperature scales, based on 
existing kagome lattice Mott insulators.

For the quantum anomalous Hall effect we showed that if the Fermi
level is placed into the kagome flat band,
the reconstructed bands are fully spin-polarized and show a
topologically non-trivial gap at the Fermi level with conducting
surface states of only one spin species.  We demonstrated our proposal
by considering the kagome Mott insulator herbertsmithite. Although the
kagome flat band is only nearly flat in the real system, a quantum
anomalous Hall state with Curie temperature well above
$1000~\mathrm{K}$ is established upon chemical substitution. The
correlated nature of the Cu $3d$ orbitals plays an important role for
the existence of fully spin-polarized bands with a gap to the empty
minority bands.  As we have been dealing with $3d$ electrons, the
calculated spin-orbit induced band gap is of the order of
$15-20~\mathrm{meV}$.  Our scheme is applicable to $4d$ and $5d$
systems, where significantly larger spin-orbit band gaps are to be
expected, while still preserving some correlation effects.

Electron doping of herbertsmithite up to the Dirac point yields, on
the other hand, a topological insulator (QSHE). With the earlier
prediction of superconductivity close to the Dirac
point~\cite{GaHerbertsmithite}, the Ga$_x$Zn$_{1-x}$-herbertsmithite
system might present an interesting platform for building a
topological quantum computer by locally controlling the Ga:Zn ratio.

Synthesis of such doped kagome systems may be a challenge. However,
our calculations show a robust stability of the structures and
correctly predict, for instance, that the Cd-substituted
herbertsmithite distorts, as has been observed
experimentally~\cite{CdHerbertsmithite}. This gives some reassurance
about the predictive power and actual realization of the phenomena
proposed in the present work. Nevertheless, chemical doping may not be
the only route to achieve hole or electron doping in
herbertsmithite. In recent years a few alternative techniques have
been very successful in doping Mott insulators like deposition of
alkali ions~\cite{Kim2014} or gating the materials with ionic
liquids~\cite{Ye2010,Nakano2012}. For instance, it has recently become
possible to tune the critical temperature of La$_2$CuO$_{4+x}$ thin
films by gating the parent compound~\cite{Kinney2015}. Following
different doping routes may allow the realization of our predictions.

\begin{acknowledgments}
The authors thank Kira Riedl, Milan Tomi\'c, Bernd Wolf, Pascal Puphal, 
Cornelius Krellner, Hans Boschker, Jochen Mannhart, Martin Jansen, Claudia 
Felser, Flavio Yair Bruno and Alberto Rivera Calzada for fruitful discussions. 
This work was supported by the German Research Foundation (Deutsche 
Forschungsgemeinschaft) through grant SFB/TR 49. RV was supported in part by 
Kavli Institute for Theoretical Physics at the University of California, Santa 
Barbara under National Science Foundation grant No. PHY11-25915.
\end{acknowledgments}

\section*{Author Contributions}
DG and HOJ performed the calculations. RV supervised the project. All authors participated in the discussion and wrote the manuscript.
\section*{Additional information}
{\bf Competing financial interests:} The authors declare that they have no competing financial interests.

\clearpage


\section*{Supplementary material}

# Supplemental Material: Prospect of quantum anomalous Hall and quantum spin Hall effect in doped kagome lattice Mott insulators

Daniel Guterding,* Harald O. Jeschke, and Roser Valentí

*Institut für Theoretische Physik, Goethe-Universität Frankfurt,*
*Max-von-Laue-Straße 1, 60438 Frankfurt am Main, Germany*

## I. FORMATION AND DOPING ENERGIES

Total energies were evaluated using *ab-initio* density functional theory (DFT) calculations within an all-electron full-potential local orbital (FPLO) [1] basis. For the exchange-correlation functional we employ the generalized gradient approximation (GGA) [2]. Total energies were extracted from calculations converged using $8^3$ $k$-point grids.

Experimental and hypothetical crystal structures were fully relaxed using the projector augmented wave (PAW) method [3] implemented in GPAW [4] with a plane-wave cutoff of 1000 eV and the GGA exchange-correlation functional [2]. We optimized the stoichiometric structures of herbertsmithite-based compounds using $6^3$ $k$-points ($4^3$ $k$-points for non-stoichiometric structures) until forces were below $10\,\mathrm{meV}/\text{Å}$.

Hypothetical crystal structures were prepared starting from the experimental crystal structure of herbertsmithite [5], substituting zinc (Zn) atoms between the copper kagome layers by monovalent $A$=Li, Na (hole-doping) and trivalent Al, Ga, In, Sc, Y (electron-doping). We refer to these compounds as $A$-herbertsmithite, $A\mathrm{Cu}_3(\mathrm{OH})_6\mathrm{Cl}_2$ (case 1) and estimate their stability by comparison to clinoatacamite [$\mathrm{Cu}_2(\mathrm{OH})_3\mathrm{Cl}$]. We directly compare energies of $A$-substituted materials, where the dopant sits in the interlayer site versus structures where it occupied a kagome site (case 2, see Fig. 3 in main paper). To analyze the phase stability, we also evaluate total energies of $3 \times 1 \times 1$ supercell structures, where either one of the $A$-sites is vacant [$A_{0.66}\mathrm{Cu}_3(\mathrm{OH})_6\mathrm{Cl}_2$, case

3] or occupied by a Cu impurity [$A_{0.66}\mathrm{Cu}_{3.33}(\mathrm{OH})_6\mathrm{Cl}_2$, case 4].

In order to determine whether the presence of excess copper in the synthesis process makes the formation of clinoatacamite favorable compared to $A\mathrm{Cu}_3(\mathrm{OH})_6\mathrm{Cl}_2$, we compare the total energies of Cu in a copper crystal plus the energy of $A\mathrm{Cu}_3(\mathrm{OH})_6\mathrm{Cl}_2$ to the total energy of $A$ in a crystal of substance $A$ plus the total energy of clinoatacamite. Proper normalization of total energies to formula units is taken into account. The calculated energy difference determines whether the formation of metallic patches of substance $A$ is energetically favourable compared to forming $A$-substituted herbertsmithite. The formation energy for this case is defined as

$$E_{\mathrm{form}} = E_{\mathrm{Cu}} + E_{\mathrm{AH}} - (E_A + E_{\mathrm{C}}),\qquad(1)$$

where $E_{\mathrm{A}}$ refers to the pure crystalline metal $A$, $E_{\mathrm{AH}}$ to $A$-herbertsmithite and $E_{\mathrm{C}}$ to clinoatacamite. As all total energies are negative, a negative formation energy signals stability of the target compound $A\mathrm{Cu}_3(\mathrm{OH})_6\mathrm{Cl}_2$. As a consistency check, we also prepared a structure with copper on the interlayer site so that the chemical formula is identical to clinoatacamite. For this structure, we find a positive formation energy. Therefore, hypothetical Cu-herbertsmithite is unstable with respect to the formation of clinoatacamite, as expected.

For the formation energy of impurity or vacancy structures, we use

$$E_{\mathrm{form}} = y \cdot E_{\mathrm{Cu}} + E_{\mathrm{AH}} - (x \cdot E_A + E_{\mathrm{AHM}}),\qquad(2)$$

where $E_{\mathrm{AHM}}$ is the total energy of the compound $A_{1-x}\mathrm{Cu}_{3+y}(\mathrm{OH})_6\mathrm{Cl}_2$.

Formation energies for modifications of herbertsmithite are shown in Table I. As all calculated energy differences are negative, formation of $A$-herbertsmithite is always favorable compared to formation of clinoatacamite. All proposed modifications are robust against formation of vacancies or copper impurities on the (interlayer) $A$-site.

The doping energies shown in Fig. 3 of the main paper are computed from energy differences between structures with dopant atoms occupying the interlayer sites and structures with dopant atoms on a kagome site (case 2). The values for the doping energies are given in Table I, together with ionic radii taken from Ref. [6].

Having established the stability of stoichiometric compounds, we investigate the solid solution of Zn and Ga

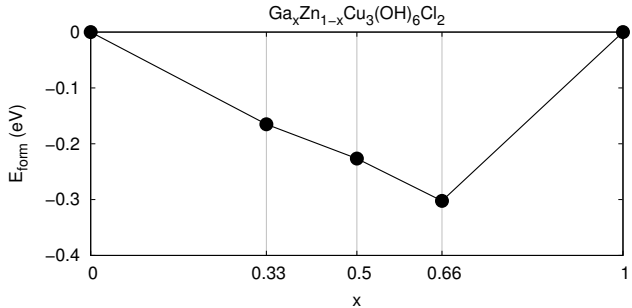

FIG. 1. Formation energies of the $\mathrm{Ga}_x\mathrm{Zn}_{1-x}\mathrm{Cu}_3(\mathrm{OH})_6\mathrm{Cl}_2$ doping series. Lines are guides to the eye.

---

* guterding@itp.uni-frankfurt.de

TABLE I. Calculated formation and doping energies for substituted herbertsmithite. Energies are given in eV (electron volts) per formula unit of $A$-substituted herbertsmithite $A\mathrm{Cu}_3(\mathrm{OH})_6\mathrm{Cl}_2$ ($A =$ Li, Na, Zn, Cd, Sc, Y, Al, Ga, In). Negative values signal stability of the stoichiometric compound against the formation of clinoatacamite (case 1), doping into the kagome layer (case 2, see Fig. 3 in main paper), vacancies on the interlayer site (case 3) and impurities on the interlayer site (case 4). Ionic radii in coordination number 6 are taken from Ref. [6]. In this configuration $\mathrm{Cu}^{2+}$ has an ionic radius of 72 pm.

| case | $A$-herbertsmithite, $A =$ | Li$^{1+}$ | Na$^{1+}$ | Zn$^{2+}$ | Cd$^{2+}$ | Sc$^{3+}$ | Y$^{3+}$ | Al$^{3+}$ | Ga$^{3+}$ | In$^{3+}$ |
|---|---|---|---|---|---|---|---|---|---|---|
| 1 | $A\mathrm{Cu}_3(\mathrm{OH})_6\mathrm{Cl}_2$ | -2.660 | -2.517 | -2.082 | -1.461 | -7.300 | -7.144 | -6.257 | -2.939 | -3.080 |
| 2 | $A\mathrm{Cu}_3(\mathrm{OH})_6\mathrm{Cl}_2$ | -0.041 | +0.257 | -0.261 | +0.128 | -0.372 | +0.043 | -1.101 | -0.736 | -0.203 |
| 3 | $A_{0.66}\mathrm{Cu}_3(\mathrm{OH})_6\mathrm{Cl}_2$ | -0.847 | -0.735 | -0.694 | n/a | -2.481 | -2.398 | -2.141 | -1.176 | -1.062 |
| 4 | $A_{0.66}\mathrm{Cu}_{3.33}(\mathrm{OH})_6\mathrm{Cl}_2$ | -1.497 | -1.350 | -1.321 | n/a | -3.019 | -2.878 | -2.659 | -1.711 | -1.585 |
| | ionic radius in pm | 76 | 102 | 74 | 95 | 75 | 90 | 54 | 62 | 80 |

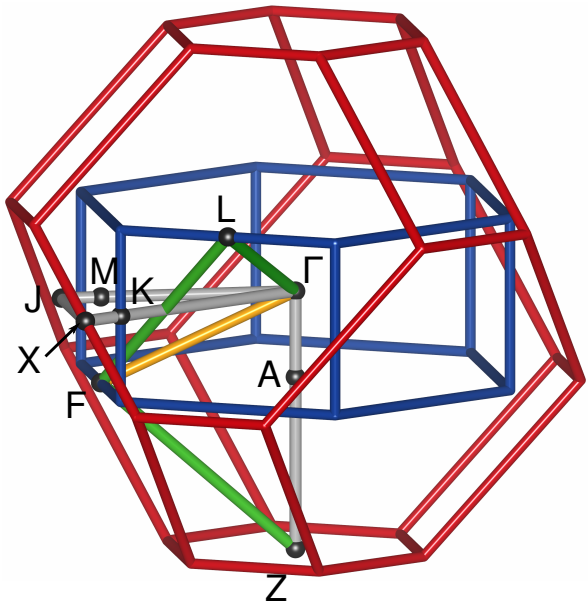

FIG. 2. Brillouin zone of herbertsmithite with high-symmetry points. The space group is $R\bar{3}m$. The path conventionally used for showing the bandstructure of herbertsmithite is indicated in grey. A path comprising the time-reversal invariant points is shown in green. The path on which surface states were calculated is shown in yellow.

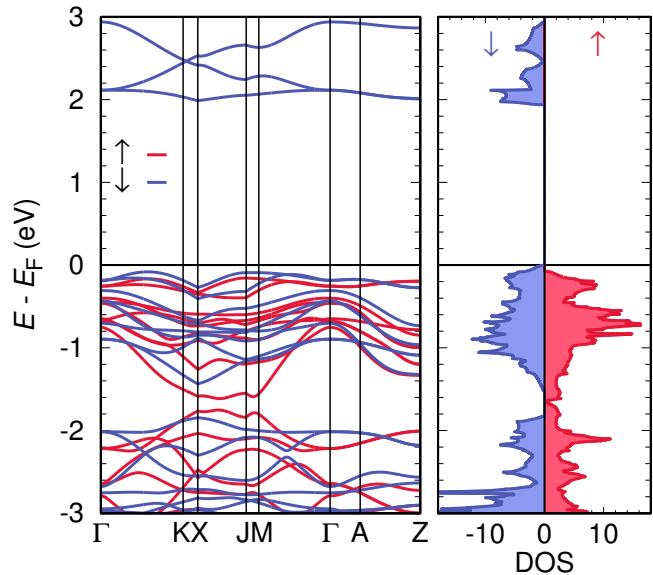

FIG. 3. Spin-polarized electronic bandstructure of herbertsmithite in ferromagnetic spin-configuration calculated from DFT+U. The interaction parameters on the Cu $3d$ shell are $U = 6$ eV and Hund's rule coupling $J_H = 1$ eV. A gap of about 2 eV is opened at the Fermi level due to electron-electron interactions.

dopants. To accomodate non-integer dopant concentrations, we use $2 \times 1 \times 1$ and $3 \times 1 \times 1$ super cells. The formation energy of the target compound having total energy $E_\mathrm{T}$ is defined as

$$E_\mathrm{form} = E_\mathrm{T} + x \cdot (E_\mathrm{Zn} - E_\mathrm{Ga}) - (1-x) \cdot E_\mathrm{ZnH} - x \cdot E_\mathrm{GaH}, \quad (3)$$

where $E_\mathrm{ZnH}$ refers to herbertsmithite and $E_\mathrm{GaH}$ to Gallium-substituted herbertsmithite. Note that as a reference energy here we use $(1-x) \cdot E_\mathrm{ZnH} + x \cdot E_\mathrm{GaH}$, which is the energy that one obtains if the solid solution were to dissociate spontaneously into herbertsmithite and Ga-substituted herbertsmithite. Therefore, formation energies are negative if the solid solution is stable against phase separation. Our results show that the solid solution $\mathrm{Ga}_x\mathrm{Zn}_{1-x}\mathrm{Cu}_3(\mathrm{OH})_6\mathrm{Cl}_2$ should exist in a broad range of doping ratios (Fig. 1).

## II. DETAILS ON PREPARATION OF HYPOTHETICAL CRYSTAL STRUCTURES

For all cases discussed in the previous section, crystal structures had to be prepared from a stoichiometric structure of herbertsmithite [5]. We started from the rhombohedral unit cell of space group $R\bar{3}m$, containing one formula unit. After reducing symmetry to space group $P1$, herbertsmithite allows for only one possibility to arrange dopant atoms or defects for all compositions investigated here, if the smallest possible supercell is considered.

Fig. 2 shows the Brillouin zone with high-symmetry points for materials based on herbertsmithite.

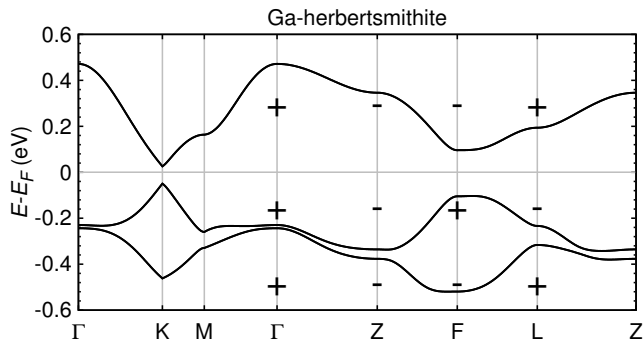

FIG. 4. Fully relativistic bandstructure of Ga-herbertsmithite. Plus and minus signs denote the parities of the three bands closest to the Fermi level at eight inversion-symmetric points ($\Gamma$, $3 \times F$, $3 \times L$, $Z$) in the Brillouin zone. In Ga-herbertsmithite the bandstructure is more three-dimensional than in Li-herbertsmithite, which leads to a substantial displacement of the Dirac point from the high-symmetry point $K$. In turn the apparent gap at $K$ is exaggerated.

## III. BANDSTRUCTURE OF HERBERTSMITHITE IN FERROMAGNETIC SPIN-CONFIGURATION

Herbertsmithite is an antiferromagnetic Mott insulator with possible quantum spin liquid ground state. Although density functional theory cannot describe such a paramagnetic ground state with large fluctuating moments, a DFT+U calculation for the most simple ordered magnetic configuration (ferromagnetic) already reproduces the insulating ground state of herbertsmithite with a large band gap of about 2 eV (see Fig. 3).

## IV. BANDSTRUCTURE AND TOPOLOGICAL PROPERTIES OF ELECTRON-DOPED HERBERTSMITHITE

We also analysed the bandstructure and topological properties of electron-doped herbertsmithite. As an example, in Fig. 4 we show the fully relativistic non-spin-polarized electronic bandstructure of Ga-herbertsmithite, where the parities of the three (dominantly Cu $d_{x^2-y^2}$) bands closest to the Fermi level are indicated. Ga-herbertsmithite is more three-dimensional than Li-herbertsmithite. Therefore, the Dirac point is displace from $K$, which exaggerates the gap at $K$. At the Dirac point the gap magnitude is the same as in Li-herbertsmithite.

The paritiy eigenvalues of the bands closest to the Fermi level are the same as in all other compounds derived from herbertsmithite that we predict to be stable. That means, topological numbers of the bands below the Dirac point are $\nu_0; (\nu_1, \nu_2, \nu_3) = 0; (111)$.

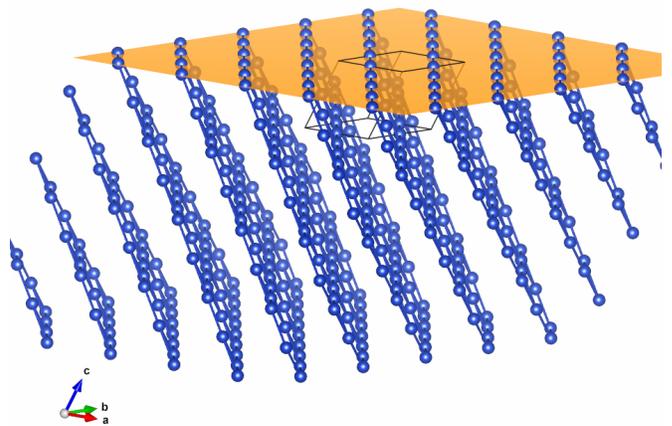

FIG. 5. (001) surface of herbertsmithite (shown as orange colored plane). Here, we used the predicted structure of Li-herbertsmithite. Ions other than Cu are omitted for clarity. The rhombohedral unit cell is shown in black. Note how all kagome layers are terminated with chains of copper ions at the surface.

## V. FORMALISM FOR COMPUTING SURFACE STATES

We use the Green's function method for the semi-infinite system [7–9] to calculate the states on the (001) surface of interlayer substituted herbertsmithite. In this scheme, the crystal must be partitioned into so-called principal layers parallel to the surface of interest. The Hamiltonian within the principal layer is denoted as $H_0$, while the coupling between neighboring principal layers is denoted as $C$. Note that only hoppings with components in direction either negative or positive perpendicular to the surface are included into $C$. The corresponding couplings in the reverse direction are taken into account in the formalism by the adjoint matrix $C^\dagger$.

The surface Green's function of a system consisting of $N$ principal layers can be written as

$$G_{ij}^{(N)}(\omega) = \left[ \left( \omega - H_0 - C G^{(N-1)} C^\dagger \right)^{-1} \right]_{ij}. \quad (4)$$

The initial condition is

$$G_{ij}^{(1)}(\omega) = (\omega - H_0)_{ij}^{-1}. \quad (5)$$

Indices $i, j$ denote combined site, orbital and spin variables. The Green's function of the dual surface can be calculated by iterating

$$G_{ij}^{(N)}(\omega) = \left[ \left( \omega - H_0 - C^\dagger G^{(N-1)} C \right)^{-1} \right]_{ij}. \quad (6)$$

As we use Green's functions on the real-frequency axis, an artificial imaginary part must be added to ensure numerical convergence. To this end, we transform $\omega \rightarrow \omega + i\eta$ with $\eta = 10^{-5}$ eV in the equations above.

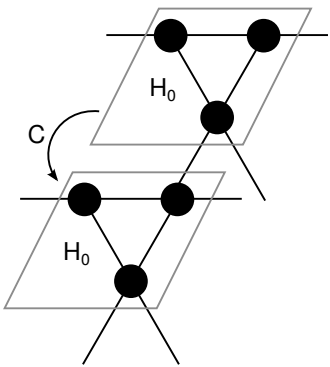

FIG. 6. Schematic partitioning of the tight-binding Hamiltonian into layers parallel to the surface. The principal layer is described by Hamiltonian $H_0$, while $C$ contains hopping elements connecting two adjacent principal layers. In the real herbertsmithite system longer range hoppings are included and two kagome layers belong to one unit cell. For comparison, see Fig. 5.

Although the method in principle allows for the treatment of semi-infinite systems, the numerical iteration of Eq. (4) or (6) has to be stopped with a finite number of layers. We use $N = 10^5$.

## VI. (001) SURFACE OF HERBERTSMITHITE

In the main paper we presented topological invariants $\nu_0; (\nu_1, \nu_2, \nu_3) = 0; (111)$ for the herbertsmithite system. According to Ref. [10], a material with those topological invariants displays conducting states on a (001) surface. The orientation of this lattice plane refers to the unit cell in rhombohedral setting of space group $R\bar{3}m$. The (001) surface of Li-herbertsmithite is shown in Fig. 5. Note how the Cu ions at the surface are arranged in chains, with all kagome layers terminated equivalently.

The unit cell of (substituted) herbertsmithite contains in total three Cu ions. These three ions comprise the minimal model of the herbertsmithite material. In the (001) surface geometry we use, two of the ions are at the surface and one is below the surface. The minimal principal layer therefore comprises Cu ions at different heights within the unit cell. This complicates the partitioning of hopping elements into matrices $H_0$ and $C$ needed for the Green's function method. The situation is depicted in Fig. 6.

As mentioned in the main paper, and obvious from Fig. 6, the semi-infinite system can be built up either in positive or negative direction perpendicular to the surface. In the kagome lattice this results in an edge termination with either chains or triangles. The case of chains is presented in the main paper. For the dual surface (triangles termination) one obtains a spectral function with equivalent features (see Fig. 7) from Eq. (6).

Both in the main paper and in the supplement we show the spectral function $A(k, \omega)$ only in positive $k$-direction

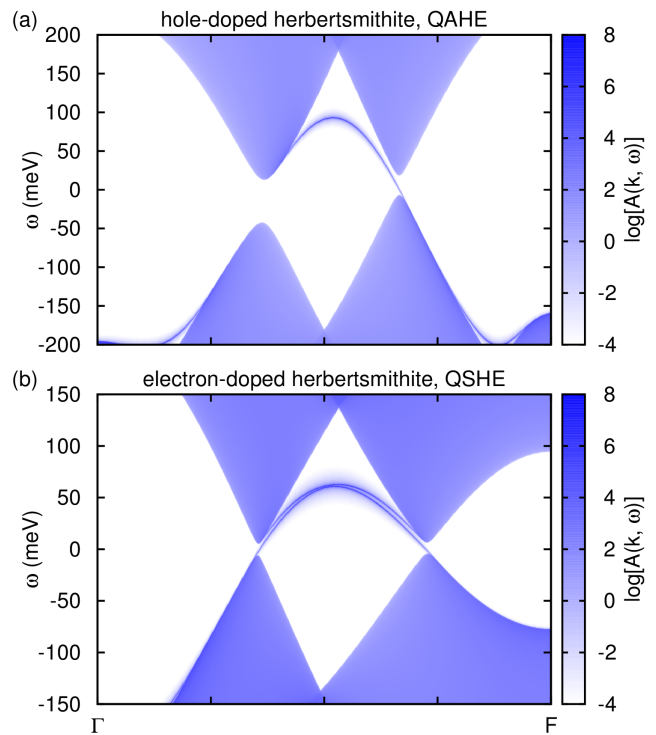

FIG. 7. Calculated states on the dual surface of substituted herbertsmithite. Spectral function on the (001) surface of (a) Li-herbertsmithite (hole-doped) and (b) Ga-herbertsmithite (electron-doped) calculated using Green's functions for the semi-infinite system. The kagome layers are terminated with triangles.

within the $k_x$-$k_y$-plane. As can bee seen from Fig. 5, the surface unit cell of (001) herbertsmithite contains two chains of Cu ions. The surface is mirror symmetric with respect to the $k_x$ and the $k_y$-direction. Therefore, the symmetry equivalent edge states of the second kagome layer appear in negative $k$-direction.

## VII. DETERMINATION OF HEISENBERG HAMILTONIAN PARAMETERS

Based on the theoretically predicted structures of Li-herbertsmithite (see Table II, Fig. 8a), we use density functional theory calculations to determine the most important couplings of the Heisenberg Hamiltonian. We use the all electron full potential local orbital (FPLO) code [1] with a generalized gradient approximation [2] exchange and correlation functional and correct for the strong correlations on the $Cu^{2+}$ $3d$ orbitals using GGA+U [11].

The results for $LiCu_3(OH)_6Cl_2$ are given in Table III. The calculation was performed with a $\sqrt{2} \times \sqrt{2} \times 2$ supercell of the rhombohedral primitive cell of Li-herbertsmithite. With $P1$ symmetry, this cell contains four formula units, $i.e.$ twelve inequivalent Cu sites. This

TABLE II. Predicted structural parameters for Li-herbertsmithite [LiCu$_3$(OH)$_6$Cl$_2$]. ($R\bar{3}m$ space group, $a = 6.96433$ Å, $c = 13.78483$ Å, $Z = 3$)

| Atom | Site | $x$ | $y$ | $z$ |
|------|------|-----|-----|-----|
| Cu | $9d$ | $^1/_2$ | $0$ | $^1/_2$ |
| Li | $3a$ | $0$ | $0$ | $0$ |
| O | $18h$ | -0.208833 | 0.208833 | 0.447033 |
| H | $18h$ | -0.131973 | 0.131973 | 0.420633 |
| Cl | $6c$ | $0$ | $0$ | 0.328170 |

TABLE III. Calculated exchange couplings for predicted Li-herbertsmithite, LiCu$_3$(OH)$_6$Cl$_2$. A GGA+U functional with $J_H = 1$ eV and the two listed $U$ values was used in combination with fully localized limit double counting correction.

| $U$ (eV) | $J_1$ (K) | $J_2$ (K) | $J_3$ (K) | $J_5$ (K) | $J_6$ (K) |
|----------|-----------|-----------|-----------|-----------|-----------|
| 6.0 | -544(29) | 7(15) | 39(24) | -50(13) | 2(8) |
| 8.0 | -332(32) | -13(16) | -11(27) | -75(20) | 1(8) |

allows for 171 out of 4096 unique spin configurations, and the data in Table III are based on fits to total energies for about 30 of these configurations. The exchange couplings are given assuming $S = 1/2$. However, due to the changed filling of the Cu bands, the moments in the calculation are reduced from 1 $\mu_B$ to, on average, about 0.67 $\mu_B$. The relevant exchange paths are visualized in Fig. 8b. Based on previous work [12], we consider the GGA+U functional with $U = 6$ eV, $J_H = 1$ eV most relevant for Cu in the square planar environment as realized in herbertsmithite type crystals.

## VIII. ESTIMATE FOR THE CURIE TEMPERATURE

We estimate the Curie temperature $T_C$ from a simple Weiss mean-field formula [13]. For $3d$ transition metals, in this approximation the Curie temperature is given by Eq. (7), where $z_i$ is the coordination number and $J_i$ are the exchange couplings in Kelvin ($U = 6$ eV), taken from Table III.

$$T_C = -\frac{2}{3}S(S+1)\sum_i z_i J_i \qquad (7)$$

We obtain $T_C = 1160$ K, where we took into account $i = \{1, 3, 5\}$ with $z_i = \{4, 4, 6\}$. Instead of the spin-1/2 model used here, it is also conceivable to parametrize the Heisenberg model with DFT effective moments for $S$, which would result in a higher value for $T_C$.

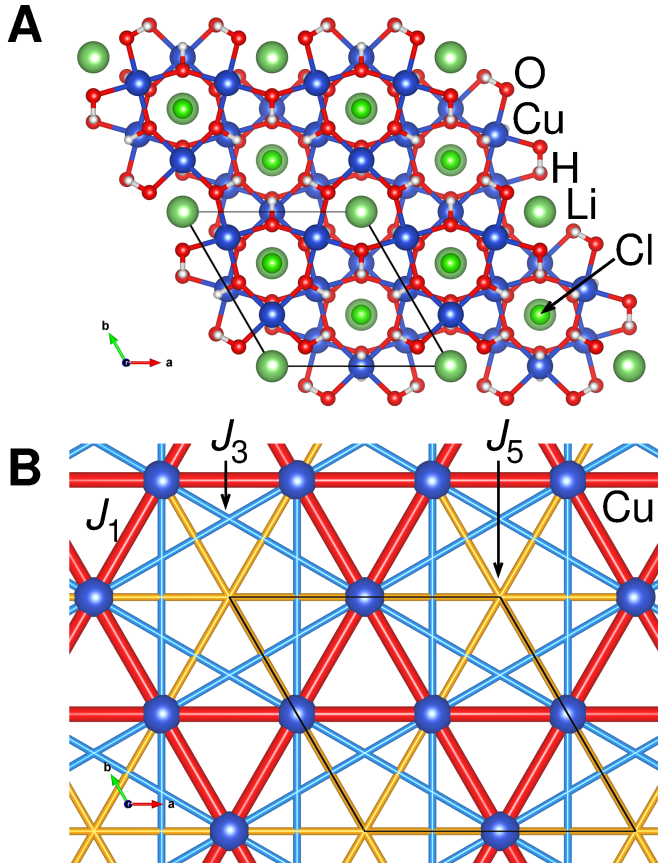

FIG. 8. Predicted crystal structure and exchange paths in Li-herbertsmithite. (a) View of the Li-herbertsmithite structure along $c$ direction (in hexagonal setting of the $R\bar{3}m$ space group). (b) Detail of the LiCu$_3$(OH)$_6$Cl$_2$ unit cell with the first three inplane exchange paths between Cu$^{2+}$ ions. Other ions are omitted for clarity.

[1] K. Koepernik and H. Eschrig, *Full-potential nonorthogonal local-orbital minimum-basis band-structure scheme*, Phys. Rev. B **59**, 1743 (1999); http://www.FPLO.de

[2] J. P. Perdew, K. Burke, and M. Ernzerhof, *Generalized Gradient Approximation Made Simple*, Phys. Rev. Lett. **77**, 3865 (1996).

[3] P. E. Blöchl, *Projector augmented-wave method*, Phys. Rev. B **50**, 17953 (1994).

[4] J. Enkovaara, C. Rostgaard, J. J. Mortensen *et al.*, *Electronic structure calculations with GPAW: a real-space implementation of the projector augmented-wave method*, J. Phys.: Condens. Matter **22**, 253202 (2010); https://wiki.fysik.dtu.dk/gpaw

[5] M. P. Shores, E. A. Nytko, B. M. Bartlett, and D. G. Nocera, *A Structurally Perfect S=1/2 Kagomé Antiferromagnet*, J. Am. Chem. Soc. **127**, 13462 (2005).




\begin{thebibliography}{99}
\bibitem{HerbertsmithiteSynthesis} Shores, M.P., Nytko, E.A., Bartlett, B.M. \& Nocera, D.G.
A Structurally Perfect S=1/2 Kagom\'e Antiferromagnet,
{\it J. Am. Chem. Soc.} {\bf 127}, 13462-13463 (2005).

\bibitem{AntisiteDisorder} Lee, S.-H. et al.
Quantum-spin-liquid states in the two-dimensional kagome antiferromagnets Zn$_x$Cu$_{4-x}$(OD)$_6$Cl$_2$.
{\it Nat. Mater.} {\bf 6}, 853-857 (2007).

\bibitem{AntisiteDisorder2} de Vries, M.A., Kamenev, K.V., Kockelmann, W.A., Sanchez-Benitez, J. \& Harrison, A.
Magnetic Ground State of an Experimental $S = 1/2$ Kagome Antiferromagnet.
{\it Phys. Rev. Lett.} {\bf 100}, 157205 (2008).

\bibitem{LeeDrought} Lee, P.A.
An End to the Drought of Quantum Spin Liquids.
{\it Science} {\bf 321}, 1306-1307 (2008).

\bibitem{BalentsQSL} Balents, L.
Spin liquids in frustrated magnets.
{\it Nature} {\bf 464}, 199-208 (2010).

\bibitem{Mendels2010} Mendels, P. \& Bert, F.
Quantum Kagome Antiferromagnet ZnCu$_3$(OH)$_6$Cl$_2$.
{\it J. Phys. Soc. Jpn.} {\bf 79}, 011001 (2010).

\bibitem{Colman2010} Colman, R.H., Sinclair, A. \& Willis, A.S.
Comparisons between Haydeeite, $\alpha$-Cu$_3$Mg(OD)$_6$Cl$_2$, and Kapellasite,
$\alpha$-Cu$_3$Zn(OD)$_6$Cl$_2$, Isostructural S = 1/2 Kagome Magnets.
{\it Chem. Mater.} {\bf 22}, 5774-5779 (2010).

\bibitem{Han2012} Han, T.-H. et al.
Fractionalized excitations in the spin-liquid state of a kagome-lattice antiferromagnet.
{\it Nature} {\bf 492}, 406-410 (2012).

\bibitem{HerbertsmithiteExchange} Jeschke H.O., Salvat-Pujol, F. \& Valent\'i, R.
First-principles determination of Heisenberg Hamiltonian parameters for the spin-1/2 kagome antiferromagnet ZnCu$_3$(OH)$_6$Cl$_2$.
{\it Phys. Rev. B} {\bf 88}, 075106 (2013).

\bibitem{Li2014} Li, Y. et al.
Gapless quantum spin liquid in the S = 1/2 anisotropic kagome antiferromagnet ZnCu$_3$(OH)$_6$SO$_4$.
{\it New J. Phys.} {\bf 16}, 093011 (2014).

\bibitem{BauerAnyons} Bauer, B. et al.
Chiral spin liquid and emergent anyons in a Kagome lattice Mott insulator.
{\it Nat. Commun.} {\bf 5}, 5137 (2014).

\bibitem{GaHerbertsmithite} Mazin, I.I. et al.
R. Thomale \& R. Valent\'i,
Theoretical prediction of a strongly correlated Dirac metal.
{\it Nat. Commun.} {\bf 5}, 4261 (2014).

\bibitem{MookEdgeStates} Mook, A., Henk, J. \& Mertig, I.
Edge states in topological magnon insulators.
{\it Phys. Rev. B} {\bf 90}, 024412 (2014).

\bibitem{PereiroSkyrmions} Pereiro, M. et al.
Topological excitations in a kagome magnet.
{\it Nat. Commun.} {\bf 5}, 4815 (2014).

\bibitem{TopologicalMagnonsChisnell} Chisnell, R. et al.
Topological Magnon Bands in a Kagome Lattice Ferromagnet.
{\it Phys. Rev. Lett.} {\bf 115}, 147201 (2015).

\bibitem{KaneMeleModel} Kane, C.L. \& Mele, E.J.
Quantum Spin Hall Effect in Graphene.
{\it Phys. Rev. Lett.} {\bf 95}, 226801 (2005).

\bibitem{BernevigZhangModel} Bernevig, B.A. \& Zhang, S.-C.
Quantum Spin Hall Effect.
{\it Phys. Rev. Lett.} {\bf 96}, 106802 (2006).

\bibitem{HasanKaneColloquium} Hasan, M.Z. \& Kane, C.L.
Colloquium: Topological insulators.
{\it Rev. Mod. Phys.} {\bf 82}, 3045-3067 (2010).

\bibitem{HaldaneModel} Haldane, F.D.M.
Model for a Quantum Hall Effect without Landau Levels: Condensed-Matter Realization of the ''Parity Anomaly''.
{\it Phys. Rev. Lett.} {\bf 61}, 2015-2018 (1988).

\bibitem{Liu2016} Liu, C.-X. Zhang, S.-C. \& Qi, X.-L.
The Quantum Anomalous Hall Effect: Theory and Experiment.
{\it Annu. Rev. Condens. Matter Phys.} {\bf 7}, 301-321 (2016).

\bibitem{KagomeQSHEWang} Wang, Z. \& Zhang, P.
Quantum spin Hall effect and spin-charge separation in a kagom\'e lattice.
{\it New J. Phys.} {\bf 12}, 043055 (2010).

\bibitem{KagomeIntrinsic} Zhang, Z.-Y.
The quantum anomalous Hall effect in kagom\'e lattices.
{\it J. Phys.: Condens. Matter} {\bf 23}, 365801 (2011).

\bibitem{FuKaneMajorana} Fu, L. \& Kane, C. L.
Superconducting Proximity Effect and Majorana Fermions at the Surface of a Topological Insulator.
{\it Phys. Rev. Lett.} {\bf 100}, 096407 (2008).

\bibitem{WilczekMajorana} Wilczek, F.
Majorana returns.
{\it Nat. Phys.} {\bf 5}, 614-618 (2009).

\bibitem{DasSarmaTopologicalComputing} Das Sarma, S. Freedman, M. \& Nayak, C.
Majorana zero modes and topological quantum computation.
{\it npj Quantum Information} {\bf 1}, 15001 (2015).

\bibitem{He2015} He, K.
The Quantum Hall Effect Gets More Practical.
{\it Physics} {\bf 8}, 41 (2015).

\bibitem{Weng2015} Weng, H. Yu, R. Hu, X. Dai, X. \& Fang, Z.
Quantum anomalous Hall effect and related topological electronic states.
{\it Adv. Phys.} {\bf 64}, 227-282 (2015).

\bibitem{QAHEMagneticTITheory} Yu, R. et al.
Quantized Anomalous Hall Effect in Magnetic Topological Insulators.
{\it Science} {\bf 329}, 61-64 (2010).

\bibitem{QAHEMagneticTIExperiment} Chang, C.-Z. et al.
Experimental Observation of the Quantum Anomalous Hall Effect in a Magnetic Topological Insulator.
{\it Science} {\bf 340}, 167-170 (2013).

\bibitem{AdatomsVanderbilt} Garrity, K.F. \& Vanderbilt, D.
Chern Insulators from Heavy Atoms on Magnetic Substrates.
{\it Phys. Rev. Lett.} {\bf 110}, 116802 (2013).

\bibitem{YanQAHEHoneycomb} Wu, S.-C. Shang, G. \& Yan, B.
Prediction of Near-Room-Temperature Quantum Anomalous Hall Effect on Honeycomb Materials.
{\it Phys. Rev. Lett.} {\bf 113}, 256401 (2014).

\bibitem{ZhangManganeseKagome} Xu, G. Lian, B. \& Zhang, S.-C.
Intrinsic Quantum Anomalous Hall effect in Kagome lattice Cs$_2$LiMn$_3$F$_{12}$.
{\it Phys. Rev. Lett.} {\bf 115}, 186802 (2015).

\bibitem{LiuOrganometallic} Wang, Z.F. Liu Z. \& Liu, F.
Quantum Anomalous Hall Effect in 2D Organic Topological Insulators.
{\it Phys. Rev. Lett.} {\bf 110}, 196801 (2013).

\bibitem{AokiOrganometallic} Yamada, M.G. et al.
First-Principles Design of a Half-Filled Flat Band of the Kagome Lattice in Two-Dimensional Metal-Organic Frameworks.
arXiv:1510.00164 (unpublished).

\bibitem{KagomeDMFT} Ohashi, T. Kawakami, N. \& Tsunetsugu, H.
Mott Transition in Kagom\'e Lattice Hubbard Model.
{\it Phys. Rev. Lett.} {\bf 97}, 066401 (2006).

\bibitem{WenHighTempFQHE} Tang, E. Mei, J.-W. \& Wen, X.-G.
High-Temperature Fractional Quantum Hall States.
{\it Phys. Rev. Lett.} {\bf 106}, 236802 (2011).

\bibitem{Hanisch1997} Hanisch, T. Uhrig, G. \& M\"uller-Hartmann, E.
Lattice dependence of saturated ferromagnetism in the Hubbard model.
{\it Phys. Rev. B} {\bf 56}, 13960-13982 (1997).

\bibitem{PAWmethod} Bl\"ochl, P.E.
Projector augmented-wave method.
{\it Phys. Rev. B} {\bf 50}, 17953-17979 (1994).

\bibitem{GPAWmethod} Enkovaara, J. et al.
Electronic structure calculations with GPAW: a real-space implementation of the projector augmented-wave method.
{\it J. Phys.: Condens. Matter} {\bf 22}, 253202 (2010).

\bibitem{PerdewBurkeErnzerhof} Perdew, J.P. Burke, K. \& Ernzerhof, M.
Generalized Gradient Approximation Made Simple.
{\it Phys. Rev. Lett.} {\bf 77}, 3865-3868 (1996).

\bibitem{FPLOmethod} Koepernik, K. \& Eschrig, H.
Full-potential nonorthogonal local-orbital minimum-basis band-structure scheme.
{\it Phys. Rev. B} {\bf 59}, 1743-1757 (1999).

\bibitem{Liechtenstein95} Liechtenstein, A.I., Anisimov, V.I. \& Zaanen, J.
Density-functional theory and strong interactions: Orbital ordering in Mott-Hubbard insulators.
{\it Phys. Rev. B} {\bf 52}, R5467-R5470 (1995).

\bibitem{FPLOtightbinding} Eschrig, H. \& Koepernik, K.
Tight-binding models for the iron-based superconductors.
{\it Phys. Rev. B} {\bf 80}, 104503 (2009).

\bibitem{SemiInfiniteGreensFunctionMethod1} L\'opez Sancho, M.P. L\'opez Sancho, J.M. \& Rubio, J.
Quick iterative scheme for the calculation of transfer matrices: application to Mo(100).
{\it J. Phys. F: Met. Phys.} {\bf 14}, 1205-1215 (1984).

\bibitem{SemiInfiniteGreensFunctionMethod2} L\'opez Sancho, M.P., L\'opez Sancho, J.M. \& Rubio, J.
Highly convergent schemes for the calculation of bulk and surface Green functions.
{\it J. Phys. F: Met. Phys.} {\bf 15}, 851-858 (1985).

\bibitem{SemiInfiniteGreensFunctionMethod3} Dai, X. Hughes, T.L., Qi, X.-L., Fang, Z. \& Zhang, S.-C.
Helical edge and surface states in HgTe quantum wells and bulk insulators.
{\it Phys. Rev. B} {\bf 77}, 125319 (2008).


\bibitem{IonicRadii} Shannon, R.D.
Revised Effective Ionic Radii and Systematic Studies of Interatomic Dinstances in Halides and Chalcogenides.
{\it Acta Cryst. A} {\bf 32}, 751-767 (1976).

\bibitem{BlundellMagnetism} Blundell, S.
{\it Magnetism in Condensed Matter}
(Oxford University Press, Oxford, 2010).


\bibitem{TIInvariants} Fu, L. \& Kane, C.L.
Topological insulators with inversion symmetry.
{\it Phys. Rev. B} {\bf 76}, 045302 (2007).

\bibitem{TMHalide} Zhou, L. et al.
New Family of Quantum Spin Hall Insulators in Two-dimensional Transition-Metal Halide with Large Nontrivial Band Gaps.
{\it Nano Lett.} {\bf 15}, 7867-7872 (2015).

\bibitem{Han2014} Han, T.-H., Singleton, J. \& Schlueter, J.A.
Barlowite: A Spin-1/2 Antiferromagnet with a Geometrically Perfect Kagome Motif.
{\it Phys. Rev. Lett.} {\bf 113}, 227203 (2014).

\bibitem{BarlowiteTheoryExperiment} Jeschke, H.O. et al.
Barlowite as a canted antiferromagnet: Theory and experiment.
{\it Phys. Rev. B} {\bf 92}, 094417 (2015).

\bibitem{CdHerbertsmithite} McQueen, T.M. et al.
CdCu$_3$(OH)$_6$Cl$_2$: A new layered hydroxide chloride.
{\it J. Solid State Chem.} {\bf 184}, 3319-3323 (2011).

\bibitem{Kim2014} Kim, Y.K. et al.
Fermi arcs in a doped pseudospin-1/2 Heisenberg antiferromagnet.
{\it Science} {\bf 345}, 187-190 (2014).

\bibitem{Ye2010} Ye, J.T. et al.
Liquid-gated interface superconductivity on an atomically flat film.
{\it Nature Mater.} {\bf 9}, 125-128 (2010).

\bibitem{Nakano2012} Nakano, M. et al.
Collective bulk carrier delocalization driven by electrostatic surface charge accumulation.
{\it Nature} {\bf 487}, 459-462 (2012).

\bibitem{Kinney2015} Kinney, J., Garcia-Barriocanal, J. \& Goldman, A.M.
Homes scaling in ionic liquid gated La$_2$CuO$_{4+x}$ thin film, 
{\it Phys. Rev. B} {\bf 92}, 100505(R) (2015).
\end{thebibliography}
\end{document}